\documentclass[preprintnumbers,prd,twocolumn,showpacs,floatfix,preprintnumbers,letterpaper,amsmath,amssymb,nofootinbib,superscriptaddress]{revtex4}
\usepackage{graphicx}
\usepackage{epsfig}
\usepackage{bm}
\usepackage{amsfonts}

\usepackage{color}% use if color is used in text

\newbox\pippobox

\newcommand {\lla} {\ {\raise-.5ex\hbox{$\buildrel<\over\sim$}}\ }

\usepackage[T1]{fontenc}
\usepackage[latin1]{inputenc}
\usepackage{graphicx}
\usepackage[english]{babel}
\usepackage{amsmath}
\usepackage{amssymb}
\usepackage{amsfonts}

\begin{document}

\title{Cyclic cosmology from Lagrange-multiplier modified gravity}

\author{Yi-Fu Cai }
\email{ycai21@asu.edu} \affiliation{ Department of Physics, Arizona
State University, Tempe, AZ 85287, USA} \affiliation{ Institute of
High Energy Physics, Chinese Academy of Sciences, P.O. Box 918-4,
Beijing 100049, China}

\author{Emmanuel N. Saridakis}
\email{msaridak@phys.uoa.gr} \affiliation{College of Mathematics
and Physics,\\ Chongqing University of Posts and
Telecommunications, Chongqing 400065, China }

\begin{abstract}
We investigate cyclic and singularity-free evolutions in a
universe governed by Lagrange-multiplier modified gravity, either
in scalar-field cosmology, as well as in $f(R)$ one. In the scalar
case, cyclicity can be induced by a suitably reconstructed simple
potential, and the matter content of the universe can be
successfully incorporated. In the case of $f(R)$-gravity,
cyclicity can be induced by a suitable reconstructed second
function $f_2(R)$ of a very simple form, however the matter
evolution cannot be analytically handled. Furthermore, we study
the evolution of cosmological perturbations for the two scenarios.
For the scalar case the system possesses no wavelike modes due to
a dust-like sound speed, while for the $f(R)$ case there exist an
oscillation mode of perturbations which indicates a dynamical
degree of freedom. Both scenarios allow for stable parameter
spaces of cosmological perturbations through the bouncing point.
\end{abstract}


\maketitle

\section{Introduction}

Inflation is now considered to be a crucial part of the
cosmological history of the universe \cite{inflation}, however the
so called ``standard model of the universe'' still faces the
problem of the initial singularity. Such a singularity is
unavoidable if inflation is realized using a scalar field while
the background spacetime is described by the standard Einstein
action \cite{Borde:1993xh}. As a consequence, there has been a lot
of effort in resolving this problem through quantum gravity
effects or effective field theory techniques.

A potential solution to the cosmological singularity problem may
be provided by non-singular bouncing cosmologies
\cite{Mukhanov:1991zn}. Such scenarios have been constructed
through various approaches to modified gravity, such as the
Pre-Big-Bang \cite{Veneziano:1991ek} and the Ekpyrotic
 \cite{Khoury:2001wf} models, gravity actions with higher order
corrections \cite{Brustein:1997cv}, braneworld scenarios
\cite{Shtanov:2002mb}, non-relativistic gravity \cite{Cai:2009in},
loop quantum cosmology \cite{Bojowald:2001xe} or in the frame of a
closed universe \cite{Martin:2003sf}. Non-singular bounces may be
alternatively investigated using effective field theory
techniques, introducing matter fields violating the null energy
condition \cite{Cai:2007qw} leading to observable predictions too
\cite{Cai:2008ed}, or introduce non-conventional mixing terms
\cite{Saridakis:2009jq}. The extension of all the above bouncing
scenarios is the (old) paradigm of cyclic cosmology \cite{tolman},
in which the universe experiences the periodic sequence of
contractions and expansions, which has been rewaked the last years
\cite{Steinhardt:2001st} since it brings different insights for
the origin of the observable universe
\cite{Lidsey:2004ef,Xiong:2008ic,cyclic} (see
\cite{Novello:2008ra} for a review).

On the other hand, the interesting idea of modifying gravity using
Lagrange multipliers, first introduced in \cite{Mukhanov:1991zn}
with bouncing solution studied in \cite{Brandenberger:1993ef}, was
recently gained new interest, since its extended version may
reduce irrelevant degrees of freedom in modified gravity models
\cite{Lim:2010yk,Gao:2010gj,Capozziello:2010uv}. In particular,
using two scalar fields, one of which is the Lagrange multiplier,
leads to a constraint of special form on the second scalar field,
and as a result the whole system contains a single dynamical
degree of freedom.

In the present work we are interested in constructing a scenario
of cyclic cosmology in universe governed by modified gravity with
Lagrange multipliers. We do that both in the case of scalar
cosmology, as well as in the case of $f(R)$-gravity, both in their
Lagrange-multiplier modified versions. As we show, the realization
of cyclicity and the avoidance of singularities is
straightforward.

This paper is organized as follows. In section \ref{modgrav} we
present modified gravity with Lagrange multipliers, separately for
scalar cosmology (subsection \ref{scalarmodel}) and for
$f(R)$-gravity (subsection \ref{frmodgrav}). In section
\ref{cyclicity} we construct the scenario of cyclicity
realization, reconstructing the required potential (subsection
\ref{scalarcycl}) and the form of $f(R)$ modification (subsection
\ref{frcyclic}). In section \ref{stability} we investigate the
stability of the cosmological perturbations of the two scenarios.
Finally, section \ref{conclusions} is devoted to the summary of
our results.

\section{Modified gravity with Lagrange multipliers}
\label{modgrav}

Let us present the cosmological scenarios with Lagrange
multipliers. In order to be general and complete, we present both
the scalar case, as well as the $f(R)$-gravity one. Throughout the
work we consider a flat Friedmann-Robertson-Walker geometry with
metric
\begin{equation}\label{metric}
 ds^{2} = -dt^{2}+a^{2}(t)d\textbf{x}^2,
\end{equation}
where $a$ is the scale factor, although we could straightforwardly
generalize our results to the non-flat case too.

\subsection{Scalar cosmology with Lagrange multiplier}
\label{scalarmodel}

In this subsection, we consider a scenario with two scalars,
namely $\phi$ and $\lambda$, where the second scalar is a Lagrange
multiplier which constrains the field equation of the first one.
The action reads \cite{Lim:2010yk, Capozziello:2010uv}:
\begin{eqnarray}
\label{LagS1}
&& S = \int d^4 x \sqrt{-g} \left\{
\frac{R}{2\kappa^2} - \frac{\omega(\phi)}{2} \partial_\mu \phi
\partial^\mu \phi - V(\phi)\right.\nonumber\\
&&\left.\ \ \ \ \ \ \ \ \ \ \ \  \ \ \ \ \ \ \ - \lambda \left[
\frac{1}{2}
\partial_\mu \phi
\partial^\mu \phi + U(\phi)\right]+\mathcal{L}_m  \right\},
 \end{eqnarray}
  where $\lambda$ is the
Lagrange multiplier field. Furthermore, $V(\phi)$ and $U(\phi)$
are potentials of $\phi$, while the function $\omega(\phi)$ and
especially its sign determines the nature of the scalar field
$\phi$, that is if it is canonical or phantom. Finally, the term
$\mathcal{L}_m $ accounts for the matter content of the universe.

Variation over the $\lambda$-field leads to
\begin{equation}
 \label{constr1}
  0 = \frac{{\dot\phi}^2}{2} - U(\phi).
\end{equation} As we observe, this equation acts as a
constraint for the other scalar field $\phi$, and it is the cause
of the significant cosmological implications of such a
construction. Now, the Friedmann equations straightforwardly write
\begin{eqnarray}
\label{Fr1}
 \frac{3}{\kappa^2} H^2 &=& \left[\frac{\omega(\phi) +
\lambda}{2}\right]{\dot\phi}^2 + V(\phi) + \lambda
U(\phi)\nonumber\\
& =& \left[ \omega(\phi) + 2\lambda \right] U(\phi) + V(\phi)
+\rho_m,\end{eqnarray}
\begin{eqnarray}
 \label{Fr2}
- \frac{1}{\kappa^2} \left(2 \dot H + 3H^2 \right) &=&
\left[\frac{\omega(\phi) + \lambda}{2}\right]{\dot\phi}^2 -
V(\phi) - \lambda U(\phi)\nonumber\\ &=& \omega(\phi) U(\phi) -
V(\phi)+p_m,
\end{eqnarray}
where we have defined the Hubble parameter as $H\equiv\frac{\dot
a}{a}$, and we have also made use of the constraint
(\ref{constr1}). Additionally, in these expressions $\rho_m$ and
$p_m$ stand respectively for the energy density and pressure of
matter, with equation-of-state parameter $w_m\equiv p_m/\rho_m$.
Observing the above equations, we can straightforwardly define the
energy density and pressure for the dark-energy sector as
\begin{eqnarray}\label{rhoDE}
&&\rho_{de}\equiv \left[ \omega(\phi) + 2\lambda \right] U(\phi) +
V(\phi) ,\\
 &&\label{pDE} p_{de}\equiv \omega(\phi) U(\phi) - V(\phi),
\end{eqnarray}
and thus the dark energy equation-of-state parameter will be
\begin{equation}
 \label{Wde}
  w_{de}
= \frac{\omega(\phi) U(\phi) - V(\phi)}{\left[ \omega(\phi) +
2\lambda \right] U(\phi) + V(\phi)} .
 \end{equation}

Let us now explore some general features of the scenario at hand
following \cite{Capozziello:2010uv}. First of all the constraint
equation (\ref{constr1}) can be integrated for positive $U(\phi)$,
leading to
\begin{equation}
 \label{tphi}
  t = \pm \int^\phi \frac{d\phi}{\sqrt{2U(\phi)}}.
\end{equation}
Inverting this relation with respect to $\phi$, one can find the
explicit $t$-dependence of $\phi$, that is the corresponding
$\phi(t)$. Thus, substituting the expression of $\phi(t)$ into
(\ref{Fr2}), we obtain a differential equation for $H(t)$:
\begin{eqnarray}
 \label{diffHt}
&&2 \dot H(t) + 3[H(t)]^2  =-\kappa^2\left\{\omega(\phi(t))
U(\phi(t))\right. \nonumber\\
&&\left.\ \ \ \ \ \ \ \ \ \ \ \ \ \ \  \ \ \ \ \ \ \ \ \ \ \ -
V(\phi(t))+w_m\rho_m(t)\right\},
\end{eqnarray}
the solution of which determines completely the cosmological
evolution. Finally, substituting $\phi(t)$ and $H(t)$ into
(\ref{Fr1}), we can find the $t$-dependence of the Lagrange
multiplier field $\lambda$:
\begin{eqnarray}
\label{lambdat}
 &&\lambda(t) =
\frac{1}{2U\left(\phi\left(t\right)\right)} \left\{
\frac{3}{\kappa^2} H(t)^2 -
V\left(\phi\left(t\right)\right)-\rho_m(t)\right\} \ \ \ \nonumber\\
&&\ \ \ \ \ \ \ \ \ \ \ \ \ \ \  \ \ \ \ \ \ \ \ \ \ \  \ \ \ \ \
\ \ \ \ \ \ \ \ \ \ \  \ \ \ \ \ -
\frac{\omega\left(\phi\left(t\right)\right)}{2}.
 \end{eqnarray}

However, one could alternatively follow the inverse procedure,
that is to first determine the specific behavior of $H(t)$ and try
to reconstruct the corresponding potential $V(\phi)$, which is
responsible for the $\phi$-evolution that leads to such a $H(t)$.
In particular, he first chooses a suitable $U(\phi)$ which will
lead to an easy integration of (\ref{tphi}), allowing for an
interchangeable  use of $\phi$ and $t$ through
$\phi(t)\leftrightarrow t(\phi)$. Therefore, the second Friedmann
equation (\ref{Fr2}) gives straightforwardly
\begin{eqnarray}
 \label{Vphit}
  &&V(\phi) = \frac{1}{\kappa^2}
\left[2 \dot H\left(t\left(\phi\right)\right) + 3 H
\left(t\left(\phi\right)\right)^2 \right] \ \ \ \ \ \ \ \ \ \  \ \ \ \ \ \ \ \ \ \ \nonumber\\
&&\ \ \ \ \ \ \ \ \ \ \ \ \ \ \ \  \  \ \ \ \ \ \  \ \ \ \ \
 + \omega(\phi)
U(\phi)+w_m\rho_m(t(\phi)),
   \end{eqnarray}
where $\omega(\phi)$ can still be arbitrary. In summary, such a
potential $V(\phi)$ produces a cosmological evolution with the
chosen $H(t)$.

Finally, let us make a comment  here concerning the nature of
models with a second, Lagrange-multiplier field, since it is
obvious that the constraint (\ref{constr1}) changes completely the
dynamics,  comparing to  the conventional models. In principle,
one expects to have a propagation mode for each new field.
However, due to the constraint (\ref{constr1}) and the form in
which the $\lambda$-field appears in the action, the propagating
modes of $\phi$ and $\lambda$ do not appear, and the system is
driven by a system of two first-order ordinary differential
equations, one for each field. As a consequence, there are no
propagating wave-like degrees of freedom and the sound speed for
perturbations is exactly zero irrespective of the background
solution  \cite{Lim:2010yk}.

\subsection{$f(R)$-gravity with Lagrange multiplier}
\label{frmodgrav}

In this subsection we present $f(R)$-gravity with Lagrange
multipliers, following \cite{Capozziello:2010uv}. In this case, we
start by a conventional $f(R)$-gravity, and we add a scalar field
$\lambda$ which is a Lagrange multiplier leading to a constraint.
In particular, the action reads:
 \begin{equation}
\label{Lag1fr}
 S = \int d^4 x \sqrt{-g} \left\{ f_1(R) - \lambda
\left[ \frac{1}{2}
\partial_\mu R \partial^\mu R
+ f_2 (R) \right]\right\} ,
 \end{equation}
where $f_1(R)$ and $f_2(R)$ are two independent functions of the
Ricci scalar $R$. Note that in the above action we have not
included the matter content of the universe, since this would
significantly modify the multiplying terms of $\lambda$, making
the subsequent reconstruction procedures technically very
complicated. These subtleties are caused by the fact that the
Lagrange multiplier field propagates in this case, which is a
disadvantage of the present scenario, contrary to the scalar
cosmology of the previous subsection.

Variation over $\lambda$ leads to the constraint
\begin{equation}
\label{constrfr}
 0= - \frac{1}{2}{\dot R}^2 + f_2(R).
 \end{equation}
Additionally, varying over the metric $g_{\mu\nu}$ and keeping the
$(0,0)$-component we obtain
\begin{widetext}
\begin{eqnarray}
\label{frfrnew}
 0 = - \frac{1}{2} f_1(R)  + 18\lambda(\ddot{H}+4H\dot{H})^2  + \left\{3\left( \dot H + H^2 \right) - 3H\frac{d}{dt} \right\} \left\{ {f_1}_{,R} - \lambda {f_2}_{,R} + \left(\frac{d}{dt} + 3H \right) \left( \lambda \frac{dR}{dt} \right) \right\}~,
\end{eqnarray}
\end{widetext}
where $_{,R}$ denotes the derivative with respect to the Ricci scalar.

For $f_2(R)>0$, the constraint (\ref{constrfr}) can be solved as
\begin{equation}
 \label{tfr}
   t = \int^R
\frac{dR}{\sqrt{2f_2(R)}},
\end{equation}
and inverting this relation with respect to $R$ one can obtain the
explicit $t$-dependence of $R$, that is the corresponding $R(t)$.
On the other hand, the Ricci scalar is given from $R = 6 \dot{H} +
12 H^2$. Thus, inserting $R(t)$ in this relation one obtains a
differential equation in terms of $H(t)$, namely
\begin{equation}
 \label{diffeqfr}
 6 \dot{H}(t) + 12 [H(t)]^2 =R(t),
\end{equation}
the solution of which determines completely the cosmological
behavior. Finally, using the obtained  $H(t)$ and $R(t)$, equation
(\ref{frfrnew}) becomes a differential equation for the Lagrange
multiplier field $\lambda(t)$ which can be solved.

However, one could alternatively follow the inverse procedure,
that is to first determine the specific behavior of $H(t)$ and try
to reconstruct the corresponding $f_2(R)$, which is responsible
for the $R$-evolution that leads to such an $H(t)$. In particular,
with a known $H(t)$ (\ref{diffeqfr}) provides immediately $R(t)$,
which can be inverted, giving  $t=t(R)$. Thus, using the
constraint (\ref{constrfr}), the explicit form of $f_2(R)$ is
found to be
\begin{equation}
 \label{f2rrecon}
 f_2(R) = \frac{1}{2}\left.
\left(\frac{d R}{dt}\right)^2 \right|_{t=t(R)}.
 \end{equation}

It is interesting to mention that in the above discussion $f_1(R)$
is arbitrary. That is, while in conventional $f(R)$-gravity the
cosmological behavior is determined completely by $f_1(R)$, in
Lagrange-multiplier modified $f(R)$-gravity the dynamics is
determined completely by $f_2(R)$. In this case $f_1(R)$ would
become relevant only in the presence of matter, and its effect on
Newton's law \cite{Capozziello:2010uv}.

\section{Cosmological Bounce and Cyclic cosmology}
\label{cyclicity}

Having presented the cosmological models with Lagrange
multipliers, both in the scalar, as well as in the $f(R)$-gravity
case, in this section we are interested in investigating the
bounce and cyclic solutions.

In principle, in order to acquire a cosmological bounce, one has
to have a contracting phase ($H<0$), followed by an expanding
phase ($H>0$), while at the bounce point we have $H=0$. In this
whole procedure the time-derivative of the Hubble parameter must
be positive, that is $\dot{H}>0$. On the other hand, in order for
a cosmological turnaround to be realized, one has to have an
expanding phase ($H>0$) followed by a contracting phase ($H<0$),
while at the turnaround point we have $H=0$, and in this whole
procedure the time-derivative of the Hubble parameter must be
negative, that is $\dot{H}<0$. Observing the form of Friedmann
equations (\ref{Fr1}), (\ref{Fr2}), as well as of (\ref{frfrnew}),
we deduce that such a behavior can be easily obtained in
principle. In the following two subsections we proceed to the
detailed investigation in the case of scalar and $f(R)$-modified
cosmology.

\subsection{Scalar cosmology with Lagrange multiplier}
\label{scalarcycl}

In order to provide a simple realization of cyclicity in this
scenario, we start by imposing a desirable form of $H(t)$ that
corresponds to a cyclic behavior. We consider a specific, simple,
but quite general example, namely we assume a cyclic universe with
an oscillatory scale factor of the form
\begin{equation} \label{at}
a(t)=A\sin(\omega t)+a_c,
\end{equation}
where we have shifted $t$ in order to eliminate a possible
additional parameter standing for the phase\footnote{Note that if
the average of an oscillatory scale factor keeps increasing
throughout the evolution, it could yield a recurrent universe
which unifies the early time inflation and late time acceleration
\cite{Feng:2004ff, Cai:2010zw}.}. Furthermore, the non-zero
constant $a_c$ is inserted in order to eliminate any possible
singularities from the model. In such a scenario $t$ varies
between $-\infty$ and $+\infty$, and $t=0$ is just a specific
moment without any particular physical meaning. Finally, note that
the bounce occurs at $a_{B}(t)=a_c-A$. Straightforwardly we find:
\begin{eqnarray}
 \label{Ht}
&&H(t)=\frac{A\omega\cos(\omega t)}{A\sin(\omega t)+a_c}\\
\label{Hpt} && \dot{H}(t)=-\frac{A\omega^2\left[A+a_c\sin(\omega
t)\right]}{\left[A\sin(\omega t)+a_c\right]^2}.
\end{eqnarray}
Concerning the matter content of the universe we assume it to be
dust, namely $w_m\approx0$, which inserted in the
evolution-equation $\dot{\rho}_m+3H(1+w_m)\rho_m=0$ gives the
usual dust-evolution $\rho_m=\rho_{m0}/a^3$, with $\rho_{m0}$ its
value at present.

Now, we first consider $\phi$ to be a canonical field, that is we
choose $\omega(\phi) = 1$. Concerning the potential $U(\phi)$ that
will give as the solution for $\phi(t)$ according to (\ref{tphi}),
we choose a simple and easily-handled form, namely
 \begin{equation}
  \label{Uphiosc}
  U(\phi) = \frac{m^4}{2},
\end{equation}
where $m$ is a constant with mass-dimension. In this case,
(\ref{tphi}) leads to
\begin{equation}
 \label{phitosc}
 \phi = m^2 t,
 \end{equation}
having chosen the $+$ sign in (\ref{tphi}). As we have said in
subsection \ref{scalarmodel}, such a simple relation for $\phi(t)$
allows to replace $t$ by $\phi/m^2$ in all the aforementioned
relations. Therefore, substitution of (\ref{Ht}),(\ref{Hpt}) and
(\ref{Uphiosc})
 into (\ref{Vphit}), and using $\phi$ instead of $t$,  provides
the corresponding expression for $V(\phi)$ that generates such an
$H(t)$-solution:
\begin{eqnarray}
 \label{Vphitosc}
 && V(\phi) = \frac{m^4}{2}+ \frac{1}{\kappa^2}
\left\{2 \left\{-\frac{A\omega^2\left[A+a_c\sin\left(\omega
\frac{\phi}{m^2}\right)\right]}{\left[A\sin\left(\omega
\frac{\phi}{m^2}\right)+a_c\right]^2}\right\}\right.\nonumber\\
&& \left.\ \ \ \ \ \ \ \ \ \  \ \ \  \ \ \  + 3
\left\{\frac{A\omega\cos\left(\omega
\frac{\phi}{m^2}\right)}{A\sin\left(\omega
\frac{\phi}{m^2}\right)+a_c}\right\}^2 \right\}  .
   \end{eqnarray}
Note that in the case of dust matter, the reconstructed potential
does not depend on the matter energy density and its evolution.
Finally, for completeness, we present also the $\lambda(t)$
evolution, which according to
 (\ref{lambdat}) reads
\begin{eqnarray}
\label{lambdatosc} && \lambda(t) = -1 - \frac{2}{\kappa^2m^4}
\left\{-\frac{A\omega^2\left[A+a_c\sin\left(\omega
t\right)\right]}{\left[A\sin\left(\omega
t\right)+a_c\right]^2}\right\}\nonumber\\
&& \ \ \ \ \ \ \ \ \  \ \ \  \ \ -\frac{1}{m^4}
\rho_{m0}\left\{A\sin\left(\omega t\right)+a_c\right\}^{-3}  .
\end{eqnarray}

\begin{figure}[ht]
\begin{center}
\mbox{\epsfig{figure=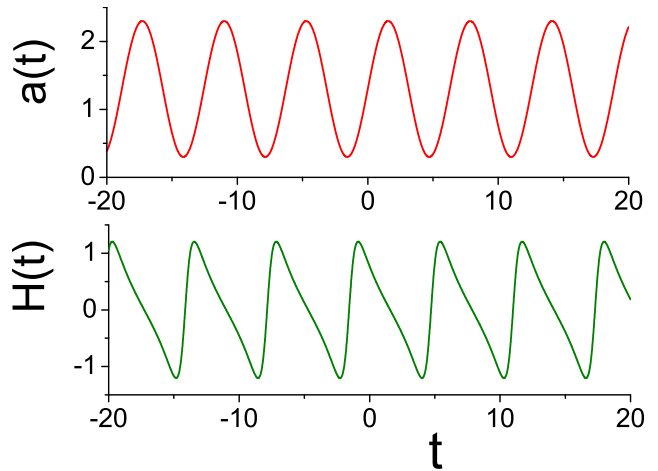,width=8.6cm,angle=0}}
\caption{{\it The evolution of the scale factor $a(t)$ and of the
Hubble parameter $H(t)$ of the ansatz (\ref{at}), with $A=1$,
$\omega=1$ and $a_c=1.3$. All quantities are measured in units
where $8\pi G=1$.
 }} \label{aHscalar}
\end{center}
\end{figure}

In order to provide a more transparent picture of the obtained
cosmological behavior, in Fig.~\ref{aHscalar} we present the
evolution of the oscillatory scale factor (\ref{at}) and of the
Hubble parameter (\ref{Ht}), with $A=1$, $\omega=1$ and $a_c=1.3$,
where all quantities are measured in units with $8\pi G=1$.
\begin{figure}[ht]
\begin{center}
\mbox{\epsfig{figure=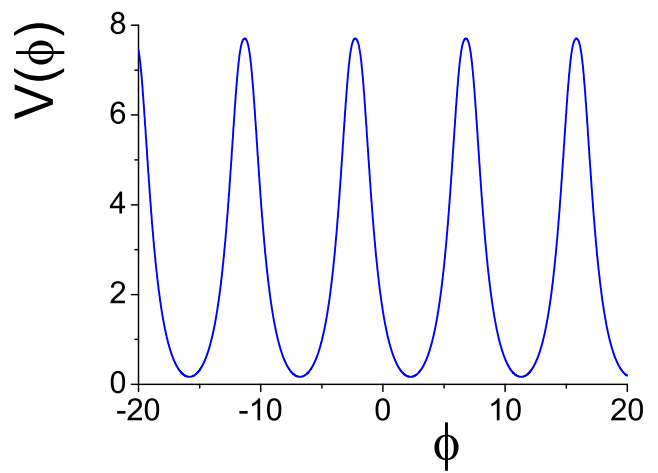,width=8.6cm,angle=0}}
\caption{{\it $V(\phi)$ from (\ref{Vphitosc}) that reproduces the
cosmological evolution of Fig.~\ref{aHscalar}, with
$\omega(\phi)=1$, $U(\phi) = m^4/2$ and $m=1.2$. All quantities
are measured in units where $8\pi G=1$.}} \label{Vscalar}
\end{center}
\end{figure}
Note that $H(t)$ by construction satisfies the requirements for
cyclicity, described in the beginning of this section.
Furthermore, in Fig.~\ref{Vscalar} we depict the corresponding
potential $V(\phi)$ given by (\ref{Vphitosc}). From these figures
we observe that an oscillating and singularity-free scale factor
can be generated by an oscillatory form of the scalar potential
$V(\phi)$ (although of not a simple function as that of $a(t)$, as
can be seen by the slightly different form of $V(\phi)$ in its
minima and its maxima). This $V(\phi)$-form was more or less
theoretically expected, since a non-oscillatory $V(\phi)$ would be
physically impossible to generate an infinitely oscillating scale
factor and a universe with a form of time-symmetry. Note also
that, having presented the basic mechanism, we can suitably choose
the oscillation frequency and amplitude in order to get a
realistic oscillation period and scale factor for the universe.
Finally, we stress that although we have presented the above
specific simple example, we can straightforwardly perform the
described procedure imposing an arbitrary oscillating ansatz for
the scale factor.

\begin{figure}[ht]
\begin{center}
\mbox{\epsfig{figure=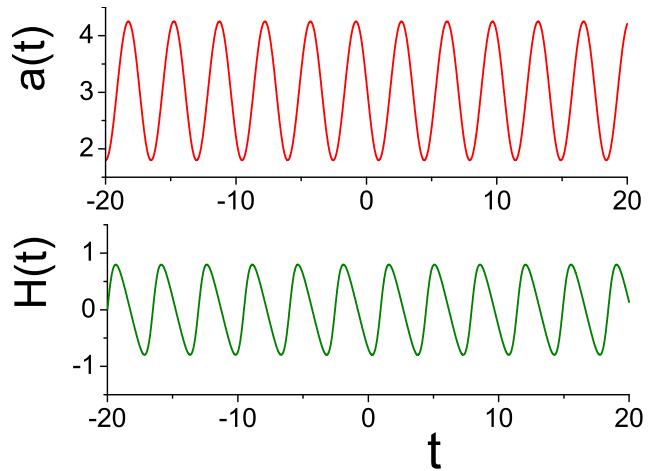,width=8.6cm,angle=0}}
\caption{{\it The evolution of the scale factor $a(t)$ and of the
Hubble parameter $H(t)$, for a scalar potential of the ansatz
(\ref{Vtspec}), with $V_0=3$, $\omega_V=1$ and $V_c=3$,
$\alpha(0)=3.2$ and $H(0)=-0.7$.  All quantities are measured in
units where $8\pi G=1$.}} \label{ahscalar2}
\end{center}
\end{figure}

The aforementioned bottom to top approach was enlightening about
the form of the scalar potential that leads to a cyclic
cosmological behavior. Therefore, one can perform the above
procedure the other way around, starting from a specific
oscillatory $V(\phi)$ and resulting to an oscillatory $a(t)$,
following the steps described in the first part of subsection
\ref{scalarmodel}. As a specific example we consider the simple
case
\begin{equation}
\label{Vtspec} V(\phi)=V_0\sin(\omega_V\, \phi)+V_c,
\end{equation}
with $U(\phi)$ chosen as in (\ref{Uphiosc}) and thus
(\ref{phitosc}) holds too. Despite the simplicity of $V(\phi)$,
the differential equation (\ref{diffHt}) cannot be solved
analytically, but it can be easily handled numerically. In
Fig.~\ref{ahscalar2} we depict the corresponding solution for
 $H(t)$ (and thus for $a(t)$) under the ansatz
(\ref{Vtspec}), with $V_0=3$, $\omega_V=1$ and $V_c=3$, with
$\alpha(0)=3.2$ and $H(0)=-0.7$ (in units where $8\pi G=1$).
As expected, we do obtain an oscillatory universe.

\subsection{$f(R)$-gravity with Lagrange multiplier}
\label{frcyclic}

In order to provide a simple realization of cyclicity in this
scenario, we start by imposing a desirable form of $H(t)$ that
corresponds to a cyclic behavior. Due to the complicated structure
of the reconstruction procedure, we will choose an easier ansatz
comparing to the previous subsection, which allows for the
extraction of analytical results. In particular, having described
the general requirements for an $H(t)$ that gives rise to cyclic
cosmology in the beginning of the present section, we choose
$H(t)$ (and not $a(t)$) to be straightaway a sinusoidal function,
that is
\begin{eqnarray}
 \label{Htfr}
H(t)=A_H\sin(\omega_H t),
\end{eqnarray}
which gives rise to a scale factor of the form
\begin{equation} \label{atfr}
a(t)=A_{H0}\,\exp\left[-\frac{A_H\cos(\omega_H
t)}{\omega_H}\right],
\end{equation}
which is oscillatory and always non-zero. Inserting (\ref{Htfr})
and its derivative into (\ref{diffeqfr}) we obtain $R(t)$ as
\begin{equation} \label{Rtfr}
R(t)=6A_H\omega_H\cos(\omega_H t)+12A_H^2\sin^2(\omega_H t),
\end{equation}
a relation that can be easily inverted giving
\begin{equation} \label{tRfr}
t(R)=\frac{1}{\omega_H}\,\arccos\left[\frac{3\omega_H+\sqrt{3(48A_H^2-4R+3\omega_H^2)}}{12A_H}\right],\nonumber
\end{equation}
where we have kept the plus sign in the square root.
 Finally,
$f_2(R)$ can be reconstructed using (\ref{f2rrecon}), leading to
\begin{eqnarray}
 \label{f2rreconexample}
&& f_2(R) =
 -\frac{1}{4}\omega_H^2\Big(48A_H^2-4R+3\omega_H^2\Big)\ \ \ \ \ \ \  \ \ \ \ \ \ \ \ \  \ \ \ \ \ \ \nonumber\\
&&\ \ \ \
\times\left[-2R+3\omega_H^2+\omega_H\sqrt{3(48A_H^2-4R+3\omega_H^2)}\right].
 \end{eqnarray}
In summary, such an ansatz for $f_2(R)$ produces the cyclic
universe with scale factor (\ref{atfr}). Note that $f_2(R)$ has a
remarkably simple form, and this is an advantage of the scenario
at hand, since in conventional $f(R)$-gravity one needs very
refined and complicated forms of $f(R)$ in order to reconstruct a
given cosmological evolution \cite{Nojiri:2006be}. However, as we
stated in the beginning of subsection \ref{frmodgrav}, in the
$f(R)$-version of Lagrange-modified gravity one cannot incorporate
the presence of matter in a convenient way that will allow for an
analytical treatment. The absence of matter evolution in a cyclic
scenario is a disadvantage, since we cannot reproduce the
evolution epochs of the universe. Therefore it would be necessary
to construct a formalism that would allow for such a matter
inclusion, similarly to the case of Lagrange-multiplier modified
scalar-field cosmology. Such a project is in preparation.

\section{Stability analysis}
\label{stability}

A central issue in all cyclic models is the stability analysis of
the cosmological perturbations along with background evolution. In
Newtonian gauge the linear metric perturbation is given by
\begin{eqnarray}\label{metricpert}
 ds^2 = - (1+2\Phi)dt^2 + a^2(t)(1-2\Psi)d\textbf{x}^2~,
\end{eqnarray}
where the variable $\Phi$ is the so-called Newtonian potential
which describes the scalar part of metric perturbation. In the
following we will study the evolution of the Newtonian potential
 near the bouncing point, for the two above scenarios respectively.

\subsection{Scalar cosmology with Lagrange multiplier}

An explicit analysis on the Newtonian potential of the scenario of
scalar cosmology with Lagrange multiplier was performed in
\cite{Lim:2010yk}. A very remarkable feature of this model is that
the system possesses no wavelike modes and thus the sound speed of
perturbations is identically zero in all backgrounds. By virtue of
this particular property, it could be possible to control the
dangerous growth of unstable modes of cosmological perturbations
in the contracting phases of the cyclic universe.

Assuming that there is no anisotropic stress in our model, then
$\Psi=\Phi$. Consequently, we only have one degree of freedom of
cosmological perturbations. According to the analysis of
\cite{Lim:2010yk}, one deduces that when the background evolution
is dominated by the scalar field $\phi$, the perturbation equation
of the Newtonian potential can be expressed as
\begin{eqnarray}
 \Phi'' +\left(1-\frac{H''}{H'}\right)\Phi' +\left(\frac{H'}{H}-\frac{H''}{H'}\right)\Phi \simeq 0,
 \label{phiddot1}
\end{eqnarray}
where the prime denotes the derivative with respect to $\ln a$.
Note that the dust-like sound speed $c_s=0$ has been applied to
eliminate the gradient term. Equation (\ref{phiddot1}) yields the
generic solution \cite{Lim:2010yk}
\begin{eqnarray}\label{Phi1}
 \Phi \simeq D(x) \left(1-\frac{H}{a}\int^a\frac{da}{H}\right) +S(x)\frac{H}{a},
\end{eqnarray}
which is applicable in all scales. Thus, in this solution there
are two modes $D(x)$ and $S(x)$, which are arbitrary functions of
spatial coordinates, and their explicit forms can be determined by
certain boundary conditions.

Observing the second term of the right-hand-side of (\ref{Phi1}),
one may concern that the Newtonian potential might be divergent on
the bouncing point, when $H=0$. Fortunately, this does not happen
in a generic bounce model. As it was shown in phenomenological
studies of generic bounce scenarios \cite{Cai:2008qw}, one can
impose an approximate parametrization of the Hubble parameter as a
linear function of the cosmic time, that is $H=\xi t$, nearby the
bouncing point $t_B=0$, with $\xi$ being a positive constant.
Doing so, in the neighborhood of a bounce in a specific cycle the
Newtonian potential can be solved as
\begin{equation}
 \Phi_B \simeq D \left[1-\sqrt{\frac{\pi}{2}\xi}te^{-\frac{\xi t^2}{2}}(e^{\frac{\xi t^2}{2}}-1)^{\frac{1}{2}}\right]
  + S \frac{\xi t}{a_B}e^{-\frac{\xi t^2}{2}},
  \label{phibb}
\end{equation}
where $a_B$ is the value of the scale factor at the bouncing
point.

Relation (\ref{phibb}) presents convergent behavior, and thus the
perturbations are able to pass through the bounce smoothly and
without any pathology. Therefore, the scenario at hand indeed
provides a satisfactory way to realize a cyclic picture without
instability on its perturbations.

\subsection{$f(R)$-gravity with Lagrange multiplier}

When we are dealing with $f(R)$-gravity with Lagrange multiplier,
we cannot use the simplifying relation $\Psi=\Phi$, even under the
assumption of zero anisotropic stress. Therefore, in order to
study the number of degrees of freedom in this case it is
convenient to expand the action into quadratic order, which can be
formally expressed as
\begin{widetext}
\begin{eqnarray}
 S^{(2)} = \int d^4x\sqrt{-g} \bigg[ \frac{{f_1}_{,RR}}{2}\delta_1R^2
  + {f_1}\left(-\frac{g_{\mu\nu}}{2}\delta{g}^{\mu\nu}\delta_1R+\delta_2R\right) %\nonumber\\
   + \left(\frac{g_{\mu\nu}}{2}\delta{g}^{\mu\nu}\right)\lambda C_2 \bigg]~,
\end{eqnarray}
\end{widetext}
where $\delta_1R$ and $\delta_2R$ denote the perturbed Ricci
scalar at first and second order respectively. Moreover, $C_2$ is
the perturbed constraint equation at first order, which form is
given by
\begin{eqnarray}
 \frac{1}{2}\delta{g}^{\mu\nu}\partial_\mu{R}\partial_\nu{R} +\partial_\mu{R}\partial^\mu\delta_1R +{f_2}_{,R}\delta_1R =0.
\end{eqnarray}

In principle, one could worry since the above action involves two
scalar modes and higher derivative terms, which could imply
instabilities. However, this is not the case since the higher
derivative terms can be fixed by the perturbed constraint
equations. In addition, the vanishing of the $(i,j)$ component of
the perturbed Einstein equation allows to eliminate one degree of
freedom. Therefore, there is still only one mode of metric
perturbation which is able to propagate freely.

In the following, we would like to focus on the propagation of the
metric perturbation around the bouncing point, which is crucial to
the stability analysis of a cyclic scenario. For calculation
convenience we assume that the bounce takes place slowly, and then
the universe approaches a static phase around the bounce
asymptotically. Under this assumption we can obtain the kinetic
terms of the perturbation equation, which take the following
approximate form:
\begin{eqnarray}
 \ddot{Q}-\frac{\nabla^2}{a^2}{Q}+\frac{f_1}{{f_1}_{,R}}Q+...=0,
\end{eqnarray}
in which $Q\equiv\Phi+\Psi$. Although this equation is far from a
complete form, one can still extract a few important issues.
Namely, the sound speed of the perturbation is unity under the
assumption of slow bounce, as it can be read from the coefficient
before the gradient term. Furthermore, we confirm that there exist
only a single degree of freedom in the $f(R)$ cosmology with
Lagrange multiplier.

We mention here that we investigated  the stability only under the
assumption of slow bounce. And the result in this case illustrates
that it is possible for the cosmological perturbation to evolve
through the bounces. However, it is necessary to point out that a
more complete and generic analysis should be performed in order to
constrain the parameter space of the scenario. Such a general
analysis is left for future investigation.

\section{Conclusions}
\label{conclusions}

In this work we investigated cyclic evolutions in a universe
governed by Lagrange-multiplier modified gravity. In order to be
more general we considered the scenario of modified gravity
through Lagrange multipliers in both scalar-field cosmology, as
well as in $f(R)$ one.

In the case of scalar cosmology, the use of a second field that is
the Lagrange multiplier of the usual scalar, leads to a rich
cosmological behavior. In particular, one can obtain an arbitrary
cyclic evolution for the scale factor, by reconstructing suitably
the scalar potential. As expected, an oscillatory scale factor is
induced by an oscillatory potential, and we were able to perform
such a procedure starting either from the scale factor or from the
potential. Additionally, the matter sector can be also
incorporated easily, and this is an advantage of the scenario,
since it allows for the successful reproduction of the thermal
history of the universe.

In the case of $f(R)$-gravity, we considered a Lagrange multiplier
field and a second form $f_2(R)$. This scenario leads also to a
rich behavior, and one can acquire cyclic cosmology by suitably
reconstructing $f_2(R)$, which proves to be of a very simple form,
contrary to the conventional $f(R)$-gravity. However, the
complexity of the model does not allow for the easy incorporation
of the matter sector, since one cannot extract analytical results,
and this is a disadvantage of the construction.

In summary, we saw that Lagrange-multiplier modified gravity may
lead to cyclic behavior very easily, and the scenario is much more
realistic in the case of scalar cosmology. Since a necessary test
of every cosmological scenario is to examine the evolution of
perturbations through the bounce \cite{Brandenberger:2009ic}, we
extended our analysis beyond the background level, in both
scenarios we considered. For the case of scalar cosmology the
perturbation behaves as a frozen mode without oscillations, since
its sound speed is vanishing. For the case of $f(R)$-gravity and
under the assumption of slow bounce, we obtain a dynamical degree
of freedom of which the sound speed is almost unity. Thus, in
conclusion, it is possible for the cosmological perturbation to
evolve through the bounces. However, a more generic perturbation
analysis, beyond the slow bounce assumption, is needed, but since
it lies beyond the scope of this work it is left for future
investigation.

\section*{Acknowledgments}

It is a pleasure to thank the anonymous referees for valuable
comments. The work of YFC is supported in part by the Arizona
State University Cosmology Initiative.

\end{document}